\documentclass[12pt]{article}

\newcommand{\be}{\begin{equation}}
\newcommand{\ee}{\end{equation}}
\newcommand{\bea}{\begin{align}}
\newcommand{\eea}{\end{align}}

\def\({\left(}
\def\){\right)}

\newcommand{\Tr}{{\rm Tr \,}}

\usepackage{color}
\usepackage{multirow}
\usepackage{graphicx}
\usepackage{hyperref}
\usepackage{amsmath}
\usepackage{amssymb}
\usepackage{amsfonts}   
\usepackage{latexsym}   
\usepackage{slashed}
\usepackage{yfonts} 
\usepackage{verbatim}
\usepackage{wrapfig}

\begin{document}

\centerline{\Large{Heterotic sigma models on $T^8$ and the Borcherds automorphic form $\Phi_{12}$}}
\bigskip
\bigskip
\centerline{Sarah M. Harrison$^1$, Shamit Kachru$^2$, Natalie M. Paquette$^2$,}
\medskip
\centerline{Roberto Volpato$^{2,3,4}$ and Max Zimet$^2$}
\bigskip
\bigskip
\centerline{$^1$ Center for the Fundamental Laws of Nature, Harvard University}
\centerline{Cambridge, MA 02138, USA}
\medskip
\centerline{$^2$ Stanford Institute for Theoretical Physics, Stanford University}
\centerline{Palo Alto, CA 94305, USA}
\medskip
\centerline{$^3$ Dipartimento di Fisica e Astronomia `Galileo Galilei',}
\centerline{Universit\`a di Padova \& INFN sez. di Padova, Italy}
\medskip
\centerline{$^4$ Theory Group, SLAC, Menlo Park, CA 94309, USA}
\bigskip
\bigskip
\begin{abstract}
We consider the spectrum of BPS states of the heterotic sigma model with
$(0,8)$ supersymmetry and $T^8$ target, as well as its second-quantized counterpart.
We show that the counting function for such states is intimately related to Borcherds'
automorphic form $\Phi_{12}$, a modular form which exhibits automorphy for $O(2,26;{\mathbb Z})$.
We comment on possible implications for Umbral moonshine and theories of AdS$_3$ gravity.
\end{abstract}

\section{Introduction}

Studies of the entropy of supersymmetric black holes in string theory have led to the discovery of
beautiful and unexpected relations between basic objects in the theory of automorphic forms,
algebraic geometry, and indices of supersymmetric sigma models.  A basic example is the
formula of Dijkgraaf, Verlinde and Verlinde \cite{DVV}, capturing the degeneracy of 1/4-BPS dyons in the $\mathcal N=4, d=4$ string theory obtained from compactifying
type II string theory on $K3 \times T^2$ or, equivalently, the heterotic string on $T^6$.  This theory has duality group
$$ SL(2,\mathbb Z)\times SO(6,22; \mathbb Z), $$
where the electric and magnetic charges are given by 28-dimensional vectors $Q,P \in \Gamma^{6,22}$, and the $SL(2,\mathbb Z)$ factor is the electric-magnetic S-duality group of the theory.

\medskip
The degeneracy of 1/4-BPS dyons, $D(Q,P)$,  with charges $(Q,P)$ is then
$$(-1)^{Q\cdot P+1}D(Q,P)= \oint d\Omega {e^{\pi i \Tr(\Omega \cdot \Lambda)} \over \Phi_{10}(\Omega)}$$
where 
$$\Omega = \left( \begin{array}{cc}
\tau & z \\
z& \sigma \end{array} \right)$$
parametrizes the Siegel upper half-space of degree 2, and 
$$\Lambda = \left( \begin{array}{cc}
Q\cdot Q & Q\cdot P \\
Q\cdot P& P\cdot P \end{array} \right)$$
encapsulates the three T-duality invariants of the black hole charges.
The beautiful function in the denominator of the integrand is the
Igusa cusp form $\Phi_{10}(\Omega)$, a weight 10 Siegel automorphic form for the modular group $SP(2,\mathbb Z)$.
For large charges this formula for the degeneracies has asymptotic growth
$$ D(Q,P) \sim e^{\pi \sqrt{Q^2P^2-(Q\cdot P)^2}},$$
reproducing the expected result for the Bekenstein-Hawking entropy of these black holes,
$$S=\pi \sqrt{Q^2P^2-(Q\cdot P)^2}.$$

\medskip
The automorphic function $\Phi_{10}$ also has a connection to elliptic genera of symmetric powers of the K3 surface, derived in \cite{DMVV} by considering the D1-D5 system on $K3 \times S^1$, which engineers 5D black holes.  The end result is an elegant formula for the generating function,
$${\phi_{10,1}(q,y) \over \Phi_{10}(p,q,y)} = \sum_{n=0}^{\infty} p^{n-1} Z_{EG}(q,y;K3^{[n]})~.$$
(Here, $\phi_{10,1}$ is a named weak Jacobi form of weight 10 and index 1; explicit expressions can be found in \cite{DMZ}). Note the correction factor $\phi_{10,1}$ spoils the automorphy of this function. In \cite{recount}, a connection between 4D and 5D black holes \cite{Andy} was used to explain the appearance of $1/\phi_{10,1}$ upon compactification and re-derive the result of \cite{DVV}.

\medskip
The purpose of this note is to report analogous formulae governing BPS state counts
 in the heterotic sigma model with $T^8$ target.  This is the model which would naturally arise on the worldsheet of heterotic string compactifications preserving half-maximal supersymmetry in two space-time dimensions.
We will focus in this note on the physics of the 2d field theory and its supersymmetry-preserving excitations,
and mostly limit any discussion of possible space-time interpretations to the concluding section.   This work was largely motivated by trying to develop an understanding of BPS counts at Niemeier points in the moduli space of
compactifications to 3 and 2 dimensions, and their possible interpretation in light of Mathieu and Umbral moonshine \cite{Eguchi:2010ej,Umbral}, in
the picture advocated in \cite{3Dumbral}.

\section{The Borcherds modular form $\Phi_{12}$}

The hero of our story will be the Borcherds modular form $\Phi_{12}$ \cite{Borcherds}.
A nice description of the relevant aspects of this form can be found in the work of Gritsenko \cite{Gritsenko},
from which we borrow heavily.

\medskip
Let $\Pi_{2,26}$ denote the (unique) even unimodular lattice of signature $(2,26)$.  The Borcherds modular
form $\Phi_{12}$ is of weight 12 with respect to $O^{+}(\Pi_{2,26})$.  It is the unique cusp form with this
property.  

\medskip
Of great interest for us will be the following explicit multiplicative lift formulae for $\Phi_{12}$.  Consider
the split 
$$\Gamma^{2,26} = \Gamma^{1,1} \oplus \Gamma^{1,1} \oplus N(-1)$$
with $N\equiv N(R)$ chosen from among the 24 even unimodular positive-definite lattices in dimension 24, i.e., the 23 Niemeier lattices and the Leech lattice.\footnote{We use the notation $N(-1)$ to denote taking the signature of $N$ to be $(0,24)$ instead of the usual $(24,0)$.}  These lattices are uniquely classified by their root systems, $R$, which are unions of simply-lace ADE root systems of the same Coxeter number, and the Leech lattice is the unique 24-dimensional even positive-definite unimodular lattice with no roots. We will denote by $h(R)$ the Coxeter number of the root system $R$ associated with $N$, where we set $h(R) =1$ when $N$ is the Leech lattice.

\medskip
To each possible choice of $N$, we can associate the refined lattice theta series
$$\theta_N(\tau,\xi) = \sum_{\lambda \in N} e^{\pi i \tau (\lambda,\lambda) + 2\pi i (\xi,\lambda)}~.$$
Here, $(v,w)$ denotes the lattice inner product of $v,w \in N$, and $\xi \in N \otimes {\mathbb C}$ can
be thought of as a complex 24-vector of `flavor' chemical potentials refining the lattice theta function.

\medskip
Then the automorphic form $\Phi_{12}(\tau,\xi,\sigma)$, defined on
$$
\{(\tau,\xi,\sigma)\in \mathbb{C}\times (N\otimes \mathbb{C})\times \mathbb{C}\mid 2\Im(\tau)\Im(\sigma)-(\Im(\xi),\Im(\xi))>0,\ \Im(\tau)>0\}\ ,
$$
 can be constructed as a multiplicative lift of a conventional Jacobi form.
Define the Fourier coefficients $f(n,\lambda)$ via
$${\theta_{N}(\tau,\xi) \over \eta^{24}(\tau)} = \sum_{n \in {\mathbb Z}, \lambda \in N} f(n,\lambda)~ q^n~ e^{2\pi i(\xi,\lambda)}~.$$
This function is a weakly holomorphic Jacobi form of weight zero and index one for the lattice $N$.
Then, $\Phi_{12}$ is given by the formula
$$\Phi_{12}(\tau,\xi,\sigma) = q^A r^{\vec{B}} p^C \prod_{\substack{n,m \in {\mathbb Z}\\ \lambda \in N \\ (n, \lambda, m)>0}} ~(1-q^n r^\lambda p^m)^{f(mn,\lambda)}$$
$$p \equiv e^{2\pi i \sigma}, ~r^\lambda = e^{2\pi i(\xi,\lambda)}.$$
In the above, we have defined $A \equiv {1 \over 24} \sum_{\lambda \in N} f(0, \lambda), \vec{B} \equiv {1 \over 2}\sum_{\lambda > 0} f(0, \lambda)\lambda \in {1 \over 2} N, C \equiv {1 \over 48} \sum_{\lambda \in N} f(0, \lambda)(\lambda, \lambda)$ \footnote{In the interpretation of $\Phi_{12}$ as a denominator for the fake Monster Lie algebra, one views $(A, \vec{B}, C)$ as a Weyl vector; see \cite{Borcherds, Gritsenko} for details and \S4 for further comments on potential applications of the algebraic structure to physics.} and used the notation $(n, \lambda, m) >0$ to mean $m>0$, or $m=0$ and $n>0$, or $m=n=0$ and $\lambda<0$. Furthermore, $\lambda>0$ (or $<0$) means that $\lambda\in N$ has positive (respectively, negative)  scalar product  with a reference vector $x\in N\otimes \mathbb{R}$. The vector $x$ must be chosen so that $(x,\lambda)\neq 0$ for all $\lambda\in N$, and different choices of $x$ are related to one each other by automorphisms in $O(\Gamma^{2,26})$.

\medskip
More precisely, the Niemeier and Leech points define cusps in the domain of definition of $\Phi_{12}$.  These
formulae should be thought of as expansions of the modular form around the cusps.

\medskip
To make contact with the earlier work \cite{3Dumbral}, it is useful to specialize the chemical potentials as follows.
Choose a fixed lattice vector $\delta \in N$.  Then we can define
$$\Theta_{N,\delta}(\tau,z) = \sum_{\lambda \in N} q^{(\lambda,\lambda) \over 2} y^{(\delta,\lambda)}$$
with $y = e^{2\pi i z}$.  
One now obtains a Jacobi form of weight 0 and index $(\delta,\delta)/2$:
$$F^{N,\delta}(\tau,z) = {\Theta_{N,\delta}(\tau, z) \over \eta^{24}(\tau)} = \sum_{n,l \in {\mathbb Z} } f_{\delta}(n,l) q^n y^l$$
where
$$f_{\delta}(n,l) = \sum_{\substack{\lambda \in N,\\(\lambda,\delta)=l}} f(n,\lambda)~.$$
These Jacobi forms, for suitable choices of $\delta$, are the BPS counting functions
discussed in \cite{3Dumbral}.  That is, they control the coefficients in the expansion of a certain ``$F^4$'' term in the low-energy effective
action of heterotic string compactification to three dimensions, when the moduli are deformed a slight distance away from a point with Niemeier symmetry 
(the enhanced symmetry point itself having singular couplings).

\medskip
A specialized form of $\Phi_{12}$ can be obtained as a lift of these BPS counting functions as well:
$$\Phi_{12}^{N,\delta}(\tau,z,\sigma) ~=~q^A y^{B} p^C \prod_{\substack{n,m,l \in {\mathbb Z}\\  (n, l, m)>0}} (1 - q^n y^l p^m)^{f_{\delta}(mn,l)}~$$
with the prefactors $A$ and $C$ as above and $B\equiv {1 \over 2}\sum_{\lambda > 0} f(0, \lambda)(\lambda,\delta)$. This object is an automorphic form on the Siegel upper half-space
$$ \{ (\tau,z,\sigma)\in \mathbb{C}^3\mid \Im(\tau)\Im(\sigma)-\Im(z)^2>0,\ \Im(\tau)>0\}
$$
 for the group $SO^+(L_{\delta})$, where $L_{\delta}$ is the lattice of
signature (2,3) and with quadratic form
$$\begin{pmatrix}
0&0&0&0&1\\
0&0&0&1&0\\
0&0&-(\delta,\delta)&0&0\\
0&1&0&0&0\\
1&0&0&0&0
\end{pmatrix}
$$

\section{$T^8$ sigma models}

\subsection{Basic connection}

The Narain moduli space \cite{Narain} of 
compactifications of heterotic strings on $T^8$ is the double coset
$$ {\cal M} =  O(8,24;{\mathbb Z}) \backslash O(8,24) \slash O(8) \times O(24) ~.$$
This structure also arises in the non-perturbative description of heterotic strings on $T^7$ \cite{Sen}.

\medskip
One can think of ${\cal M}$ as parametrizing even
unimodular lattices of signature $(8,24)$,
$\Gamma_{8,24}({\cal M})$.
The worldsheet field theory at a given point in moduli space consists of 2d bosons propagating on the 
relevant lattice.  

\medskip
Let us consider this theory on a toroidal worldsheet.
At any point in moduli space, there are 24 abelian currents of conformal dimension $(1,0)$.  One can
consider coupling these to background chemical potentials (``Wilson lines").  In this setup, the parameter
$\tau$ of \S2 can be considered as the modular parameter of the torus, while the 24 complex chemical potentials
$\xi$ should be thought of as these Wilson line degrees of freedom.  The parameter $p$ will emerge upon
second-quantization.

\medskip
Now, let us make a precise connection to the discussion of \S2.  Consider  a point in ${\cal M}$
where
$$\Gamma_{8,24} = \Gamma_{8,0} \oplus N(-1)$$
where $N$ is again one of the 24 even unimodular positive-definite lattices, and $\Gamma_{8,0}$ the $E_8$ lattice.
We will call this CFT $V_N$.

\medskip
One obtains BPS states in the heterotic sigma model by putting right-movers in their ground state and performing a trace counting
left-moving excitations. We restrict the counting to BPS states carrying no right-moving momentum, i.e. carrying zero charge with respect to the right-moving $E_8$ lattice:
$$ {\rm Tr}_{V_N} \left( (-1)^F q^{L_0} \bar q^{\bar L_0} y^{\xi^a \cdot J_a} \right)~$$
(with $J_a$ the 24 left-moving abelian currents).
A subtlety arises because of the extra right-moving zero modes of this string background; without inserting extra factors of the right-moving fermion number operator $F_R$ in the trace, the computation
will formally vanish due to the existence of fermion zero modes.  One can circumvent this vanishing by inserting powers of $F$ in the trace to soak up zero modes, as in \cite{MMS}; however, this will alter the modularity properties and spoil the connection we intend to make to $\Phi_{12}$.   Instead, we consider, following
\cite{Paquette:2016xoo}, a `twisted' computation where we trace over left-moving excitations but consider the ground states of the ${\mathbb Z}_2$ orbifold of the right-moving sector -- that is, we consider the right-movers to live in the Ramond sector of the Conway module constructed by Duncan \cite{John}.  
\footnote{This is related by modular invariance to computations with a ${\mathbb Z}_2$ insertion in the trace, which give character valued indices of the original model -- hence, we view it as a trick to extract certain BPS degeneracies of the original theory.}
The orbifold action kills the extra zero modes, but leaves the left-moving sector untouched.  The ground-state degeneracy of the right-movers 
gives an overall factor of 24, as in \cite{Paquette:2016xoo}.  Including this degeneracy, the result (for a given $N$) is
$${\rm BPS~counting~function}(\tilde V_N) \equiv {\rm Tr}_{\tilde V_N} \left( (-1)^F q^{L_0} \bar q^{\bar L_0} y^{\xi^a \cdot J_a}\right)  =  24 {\Theta_N(\tau,\xi) \over \eta^{24}(\tau)}~$$
(where $\tilde V_N$ is the twisted Hilbert space described above).
Specializing chemical potentials as before by choosing a $\delta \in N$, we see that the BPS counting function is
$F^{N,\delta}(\tau,z)$.

\medskip
We note here that the need to consider the twisted index computation to find a non-zero answer plays well with the structure of $\Phi_{12}$.  While
the heterotic compactification without the twist has a moduli space obtained by choosing left and right moving sub-lattices of $\Gamma^{8,24}$, after
twisting the $E_8$ lattice is rigidified.  Therefore, compactifying on an additional $T^2$ to compute the index, one expects a moduli space
based on $\Gamma^{2,26}$.  This makes it natural to expect a connection to an automorphic form for the Narain modular group $O(2,26;{\mathbb Z})$,
such as $\Phi_{12}$.

\subsection{Second quantization}

Now, let us consider the second-quantized version of this counting function.  To do this, consider the heterotic
sigma model based on the conformal field theory ${\rm Sym}^n(\tilde V_N)$.  ($\tilde V_N$ appears because again, one can count BPS states without additional insertions to absorb fermion zero modes by quotienting by the natural ${\mathbb Z}_2$ action on the $E_8$ lattice, before taking the symmetric product and putting right movers into their ground state). We still restrict ourselves to BPS states carrying zero charge with respect to the right-moving $U(1)$ currents.  Following the logic of Dijkgraaf, Moore, Verlinde and Verlinde \cite{DMVV},
we see that  
$${\rm log}\left(\sum_{n=0}^{\infty} p^n F^{{\rm Sym^n}(\tilde V_N)}(\tau,z) \right)~=~24 ~{\rm log} \left({\psi_{N,\delta}(\tau,z,\sigma) \over \Phi_{12}^{N,\delta}(\tau,z,\sigma)}\right)~.$$
Here, $F^{{\rm Sym^n}(\tilde V_N)}$ is the specialized BPS counting function for the CFT based on the nth symmetric product
of the (twisted) heterotic sigma model, and 
$$\psi_{N,\delta} = \pm p^{h(R)} \eta^{24}(\tau) \prod_{\alpha \in R^+} {\theta_1(\tau,z(\delta,\alpha)) \over \eta(\tau)}$$
where $R^+$ is the set of positive roots of the lattice $N$, and the sign depends on the choice of the set of positive roots.
The factor of 24 again arises from the right-moving ground state degeneracy, as in \cite{Paquette:2016xoo}.

\medskip
We see that there is a precise analogy to the findings in \cite{DVV,DMVV}: just as an (automorphic correction of) $1\over \Phi_{10}$ governs the
BPS states of the sigma models on the Hilbert scheme of K3 surfaces, (an automorphic correction of) $1\over \Phi_{12}$ governs the BPS states of the
heterotic sigma models ${\rm Sym^n}(\tilde V_N)$.  The role of the factor $\psi_{N,\delta}$ is superficially similar 
to that of the automorphic correction $\phi_{10,1}$ in the former story.  As mentioned earlier, this factor achieves an interpretation
in the 4d/5d lift \cite{recount,Andy}, and it would be interesting to give a precise similar interpretation to $\psi_{N,\delta}$.

\section{Discussion}

In this note, we've described how the (inverse of the) Borcherds modular form $\Phi_{12}$ serves as a generating function 
for the BPS state degeneracies of heterotic sigma models on $T^8$ (after suitably twisting to kill right-moving zero modes), and their symmetric products.  We conclude with
several comments and possible avenues for further development.

\medskip
\noindent
$\bullet$ In light of AdS$_3$/CFT$_2$ duality, it is natural to conjecture that these heterotic conformal field
theories, $\text{Sym}^n(V_N)$, (at least
at large central charge, i.e. large $n$) are dual to AdS$_3$ gravity theories with discrete symmetry groups corresponding to those of the associated
Niemeier lattice -- i.e. the Umbral groups and Conway's largest sporadic group.  A criterion was developed in 
\cite{EGcriterion}, using BPS degeneracies of 2d SCFTs to test for a possible large radius gravity dual.  By
modifying and checking this criterion for the case at hand, it has been found that these theories will have (at best) `stringy' gravity duals -- they
will not achieve a (parametric) separation between the inverse AdS radius and $M_{\rm string}$ as the central
charge $c \to \infty$ \cite{Nathan}.

\medskip
\noindent
$\bullet$ The 2d heterotic compactifications governed by the CFTs $V_N$ are the dimensional reductions of the
Niemeier points studied in \cite{3Dumbral} in relation to moonshine.  It is tempting to try and connect Borcherds'
modular form $\Phi_{12}$ to Umbral moonshine.  Making these thoughts precise is difficult because the notion of 
counting BPS states in gravity theories in $d \leq 3$ flat dimensions is fraught with subtlety; charged or gravitating particles have strong infra-red effects in low dimensions.  Possibly, finding
an interpretation of the present formulae 
in the setting of the supersymmetry-protected amplitudes studied in \cite{Pioline} is a route forward.

\medskip
\noindent
$\bullet$ While the models $V_N$  depend on the choice of a Niemeier lattice $N$, the function counting the second quantized BPS states is the same in all cases. Indeed, these counting functions are just Fourier expansions at different cusps of the the same automorphic function $\Phi_{12}$.  This is non-trivial: the points where the lattice $\Gamma^{8,24}$ splits into an $E_8$ lattice plus a Niemeier lattice are isolated points in the Narain moduli space of perturbative heterotic strings on $T^8$. The basic reason behind this phenomenon is that these isolated points are the different decompactification limits in the Narain moduli space of heterotic strings on $T^9$, with the condition that the Narain lattice $\Gamma^{9,25}$ splits as an orthogonal sum of $E_8$ and $\Gamma^{1,25}$. This suggests that there might be an interpretation of $\Phi_{12}$ in terms of compactification of the heterotic string to one dimension, possibly along the lines suggested in \cite{Paquette:2016xoo}.

\medskip
\noindent
$\bullet$ It is known that $1/\Phi_{10}$ is related to the square of the denominator of a generalized Kac-Moody (GKM) algebra. There is a beautiful story relating this algebra to wall-crossing of 1/4-BPS dyons in $\mathcal N=4, d=4$ string theory \cite{Cheng:2008fc}. As we mentioned earlier, $1/\Phi_{12}$ is also the denominator of a GKM algebra--the fake monster Lie algebra. It would be interesting to explore a similar story relating this algebra to the BPS states in the theories we discuss.

\medskip
\noindent
$\bullet$ The Gromov-Witten theory of $K3 \times T^2$ has recently been conjectured to be governed by the Igusa cusp form
\cite{OP}.  This follows from the role of $\Phi_{10}$ in black hole entropy counts, and string duality. Given that the heterotic string on $T^8$ is dual to type IIA on $K3 \times T^4$, it is tempting to think that the appearance of $\Phi_{12}$ in BPS counts in the present setting presages a similar role for $\Phi_{12}$
in (flavored) enumerative geometry, perhaps of $K3 \times T^4$ \cite{unpub}. 

\bigskip
\centerline{\bf{Acknowledgements}}
\medskip
We are grateful to N. Benjamin and A. Tripathy for discussions. 
S.M.H. is supported by a Harvard University Golub Fellowship in the Physical Sciences and DOE grant DE-SC0007870.
The research of S.K. was supported in part by the 
National Science Foundation under grant NSF-PHY-1316699.
N.M.P. is supported by a National Science Foundation Graduate Research Fellowship under grant DGE-114747. 
R.V. is supported by a grant from `Programma per giovani ricercatori Rita Levi Montalcini' 2013. M.Z. is supported by the Mellam Family Fellowship at the Stanford Institute for Theoretical Physics.

\end{document}